\documentclass[conference]{IEEEtran}

\usepackage{amsthm,bm,mathrsfs,amssymb}
\usepackage[cmex10]{amsmath}
\usepackage{array}
\usepackage{indentfirst}
\usepackage{algorithm}
\usepackage{algorithmic}
\usepackage{tablefootnote}
\usepackage{multirow,booktabs,threeparttable}
\usepackage{graphicx,subfigure}
\usepackage{fancyhdr}
\usepackage{color}
\usepackage{amsmath}
\usepackage{cite}
\theoremstyle{remark}

\theoremstyle{plain}
\newtheorem{remark}{Remark}
\begin{document}

%
\title{ \! Not \!Call \!Me \!Cellular \!Any \!More: \!The \!Emergence \!of \!Scaling \!Law, \!Fractal \!Patterns \!and \!Small-World  \!in \! Wireless \!Networks}
\author{\IEEEauthorblockN{Chao Yuan, Zhifeng Zhao, Rongpeng Li, Meng Li, and Honggang Zhang \\
College of Information Science and Electronic Engineering\\
Zhejiang University, Zheda Road 38, Hangzhou 310027, China\\
Email: \{yuan\_chao, zhaozf, lirongpeng, lixiaomeng, honggangzhang\}@zju.edu.cn}
}


\maketitle

\begin{abstract}
\fussy
In conventional cellular networks, for base stations (BSs) that are deployed far away from each other, it is general to assume them to be mutually independent. Nevertheless, after long-term evolution of cellular networks in various generations, this assumption no longer holds. Instead, the BSs, which seem to be gradually deployed by operators in a service-oriented manner, have embedded many fundamentally distinctive features in their locations, coverage and traffic loading. These features can be leveraged to analyze the intrinstic pattern in BSs and even human community. In this paper, according to large-scale measurement datasets, we build up a correlation model of BSs by utilizing one of the most important features, i.e., spatial traffic. Coupling with the theory of complex networks, we make further analysis on the structure and characteristics of this traffic load correlation model. Numerical results show that the degree distribution follows scale-free property. Also the datasets unveil the characteristics of fractality and small-world. Furthermore, we apply collective influence (CI) algorithm to localize the influential base stations and demonstrate that some low-degree BSs may outrank BSs with larger degree.\\
\end{abstract}
\sloppy

\IEEEpeerreviewmaketitle

\section{Introduction}
As the theory of complex networks gets increasingly developed, it has been successfully applied to understand the embedded property in a variety of real-world complex systems from various fields, such as social, ecological, biological and public transport networks \cite{Ferber2012Fractal,Song2005ARTICLES,Zhang2012Fractality,Rozenfeld2010Small}
. In spite of the significant differences in these real-world networks, several prominent properties, including scale-free (SF) pattern, small-world, and fractality, are proven to hold in common and contribute a lot to better understanding complex networks. Scale-free pattern can be depicted by a Power-law function with respect to the degree distribution, i.e. $P(k)$$\sim$$k^{-\lambda}$, where $\lambda$ is the degree exponent and the degree $k$ denotes the number of links to a node \cite{Zhang2012Fractality}. Small-world means that although the size of network $N$ (or the number of nodes) is very large, the average distance $d$ between two randomly chosen nodes is small, being well approximated by $d$$\sim$$\ln N$ \cite{Rozenfeld2010Small}. Fractality \cite{Song2005Self,Strogatz2005Complex}, which could be generally evaluated by the box-covering algorithm \cite{Song2005Self}. Specifically, fractality implies that when the size of one covering box is $L_b$, the minimum number of boxes $N_b$ required to tile the entire networks should follow $N_b$$\sim$$L_b^{-d_b}$, where $d_b$ is the fractal dimension \cite{Zhang2012Fractality}. 

On the other hand, cellular networks have been undergoing a long history of evolution and gradually accumulated unique spatial distribution pattern, as base stations (BSs) are continually deployed to provision the ever-increasing mobile traffic in hotspots accompanied by the global popularity of smart phones and tablets. Accordingly, by taking advantage of realistic traffic records from cellular networks, we can leverage the theory of complex networks to answer what is the intrinsic evolved nature in cellular networks$?$ In particular, what is the relationship between two BSs that are distant from each other$?$ In order to answer these questions, we first create a spatial traffic correlation model of BSs by regarding BSs as nodes and the traffic correlation between BSs as edges. Then, we analyze the structure and properties of this spatial traffic correlation model and derive the corresponding results in the networks. Interestingly, we discover that there exist three key properties, i.e., scale-free pattern, fractality, and small-world. It should be noted that fractality contradicts the small-world property in essence, since the former is mainly due to the repulsion between nodes of large degree (e.g. hubs) in disassortative networks while the latter is just the reverse \cite{Song2005ARTICLES}. Hence, in order to provide more evidence to prove our results, we try to validate the results from another perspective, by calculating the Pearson coefficient and the correlation profile of the spatial traffic correlation model. Then, we extract its skeleton to search for the most close pairs of BSs and the skeleton is found to be fractal as well. The definition and description of network skeleton will be given with details in the following sections.

Furthermore, after analyzing the structural properties, we try to identify influential BSs in cellular networks based on the spatial traffic correlation model. Besides the two extensively used heuristic methods focusing on node degree (i.e., high-degree and high-degree adaptive), we also leverage the collective influence (CI) \cite{morone2015influence,morone2016collective} algorithm to evaluate the influence of each BS and find that CI algorithm performs most effectively. Particularly, we extract the most influential 500 BSs sorted by CI to verify whether the high-degree BSs are more significant in our spatial traffic correlation model.

The remainder of this paper is organized as follows. In Section \uppercase\expandafter{\romannumeral2}, we briefly introduce the real measurement datasets and necessary mathematical background.
\begin{table}
\newcommand{\tabincell}[2]{\begin{tabular}{@{}#1@{}}#2\end{tabular}}
\centering
\caption{The Datasets of BSs and the Traffic Information.}
\setlength\abovecaptionskip{-5pt}
\setlength\belowcaptionskip{-5pt}
\label{tb:bss}
\begin{tabular}{l|c|c}
\toprule
\multicolumn{3}{c}{BS information}\\
\toprule
Attributes & City A & City B \\
\midrule
Network Type & 2G cellular network & 3G cellular network \\
\hline
BS Type & \tabincell{c}{1441 microcells \\ 4132 macrocells} & 2053 microcells   \\
\hline
Location & \tabincell{c}{Longitude, \\  latitude } & \tabincell{c}{Longitude, \\  latitude }  \\
\hline
No. of BSs & 5573 & 2053   \\
\toprule
\multicolumn{3}{c}{Traffic information}\\
\toprule
Traffic Resolution & One hour & Half an hour \\
\hline
Duration & 7 days & 1 day  \\
\bottomrule
\end{tabular}
\end{table}
Section \uppercase\expandafter{\romannumeral3} describes the procedures to establish the spatial traffic correlation model. Based on such a model, in Section \uppercase\expandafter{\romannumeral4}, we study the degree distribution and uses three methods to identify the influential BSs. Section \uppercase\expandafter{\romannumeral5} focuses on the structural and characteristic analyses in cellular networks. Finally, Section \uppercase\expandafter{\romannumeral6} concludes this work.\\
%


\vspace{-10pt}
\begin{figure*}
	\centering
	\subfigure[City A]
	{ \label{fig:maps:a}
		\includegraphics[width=.4\textwidth,height=.36\textwidth]{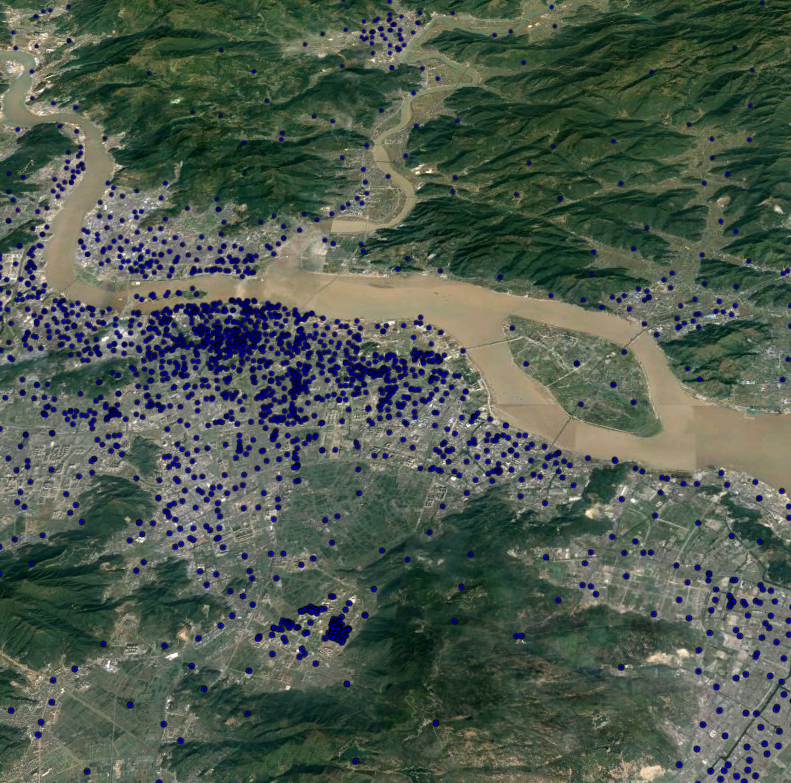}}
	\hspace{1em}
	\subfigure[City B]
	{ \label{fig:maps:b}
		\includegraphics[width=.4\textwidth,height=.36\textwidth]{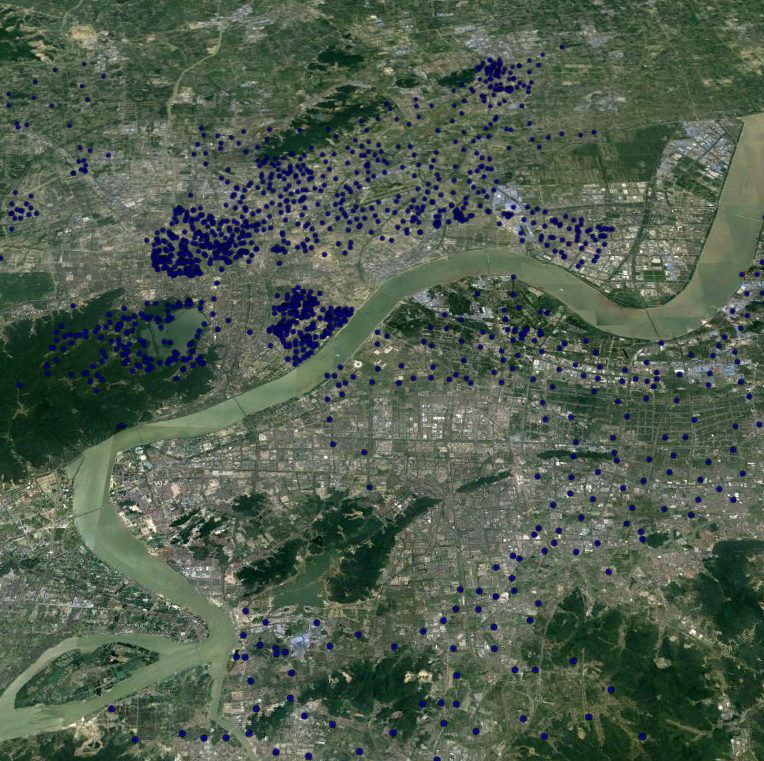}}
	\hspace{1em}
	\hspace{1in}
	\caption{An illustration of the deployment of base stations in two typical cities with geographical landforms.}
	\label{fig:maps}
\end{figure*}

\section{Background}
\subsection{Data Acquisition and Preparation}
We acquire the real measurement data from one of the biggest commercial mobile operator in China, which contains the information of traffic and BSs from a second-generation (2G) cellular network in City A and the counterpart from a third-generation (3G) cellular network in City B. Specifically, the traffic data is measured in the unit of bytes that each BS transmits to the serving users. The related traffic for City A and City B lasts 7 days and 1 day, with one-hour and half-hour granularity, respectively. Therefore, for one BS, the traffic series for City A and City B could be regarded as a vector of 168 entries and 48 entries, respectively. Meanwhile, we plot the BS deployment with the geographical landforms in Fig. \ref{fig:maps}. Moreover, the BS related information such as BS type, location area and geographic location is avaiable as well and more details are summarized in Table \uppercase\expandafter{\romannumeral1}.

\subsection{Fundamental Knowledge of Graph Theory}
Generally, the analyses of many real-world complex networks could leverage the fruits from graph theory. Without loss of generality, in graph theory, denote an undirected network as $G(V,E)$, where $V$ is the set of nodes, $E$ is the set of edges and $e_{ij} \in E$ represents the link between node $i$ and node $j$. The degree of node $i$ is defined as the number of its linked neighbours \cite{Pastorsatorras2014Epidemic}. Meanwhile, any undirected network can be described by a corresponding adjacency square matrix $W$ with the dimension of $N$, where $N$ is the size of the network. Each element $w_{ij} \in W$ equals one if there exists a link between node $i$ and node $j$, and zero otherwise \cite{Bollobas2015Modern}.

\subsection{Box-covering Algorithm}
As a widely used technique for characterizing fractal networks and calculating their fractal dimensions, box-covering algorithm has experienced a number of distinct versions since the generalized box-covering algorithm was introduced by Song $et$ $al$ \cite{Song2005Self}. The random sequential (RS) box-covering algorithm \cite{Zhang2012Fractality} is not suitable in our work due to its low efficiency in finding the minimum number of boxes among all the possible tiling configurations. Therefore, we adopt a slightly improved algorithm in \cite{Kitsak2007Betweenness} and detailed steps is shown in Algorithm. 1.

\begin{algorithm}
	\caption{Box-covering algorithm }
	\label{alg:A}
	\begin{algorithmic}
		\STATE input: undirected network : $G=(V, E)$ , adjacency matrix $W$ ;
		
		\STATE output: $S$;
		\STATE initialization:
		\STATE {set $S=\emptyset, L_b=1$.} 
		\REPEAT
		\STATE	$C=\{1,2,...,N\}$ ;
		\STATE	$N_b=0$ ;
			\REPEAT
			\STATE randomly choosen a node $i$ in $C$;
			\STATE find the set $R$ of nodes that have a distance larger than $L_b$ from node $i$;
			\STATE $N_b=N_b+1$  ;
			\STATE set $C=R$;
			\UNTIL ($C==\emptyset$)
		
		\STATE set ${S=S+\{N_b\}}$; 
		\STATE set ${L_b=L_b+1}$; 
		\UNTIL ($L_b==network \ diameter$)
	\end{algorithmic}
\end{algorithm} 

Additionally, the box number $N_b$ derived from this algorithm may not be the minimum number of the corresponding size $L_b$. In order to solve this problem, we repeat the process 1000 times and obtain 1000 values for each $L_b$, expecting that the maximum of them can approximate the desired value.

\section{Modeling Process}
\subsection{Basics}
In this part, we build undirected graph with BSs as nodes. Here, the traffic load of BS $i$ can be expressed by a traffic load vector $x_i$=[$x_i$(1), $x_i$(2), \ldots, $x_i$($T$)], where T equals 168 and 48 for City A and City B, respectively. Afterwards, we calculate Pearson correlation coefficient between any two traffic vectors and assume the corresponding results as the value. For BS $i$ and BS $j$, the Pearson correlation coefficient is defined as:

\begin{equation}
\rho_{ij} =\frac{T\sum{x_i x_j}-\sum{x_i} \sum{x_j}}{\sqrt{T\sum{x_i^2}-(\sum{x_i})^2} \sqrt{T\sum{x_j^2}-(\sum{x_j})^2}}
\label{eq30}
\end{equation}

\begin{figure}
	\centering
	\includegraphics[width=0.5\textwidth]{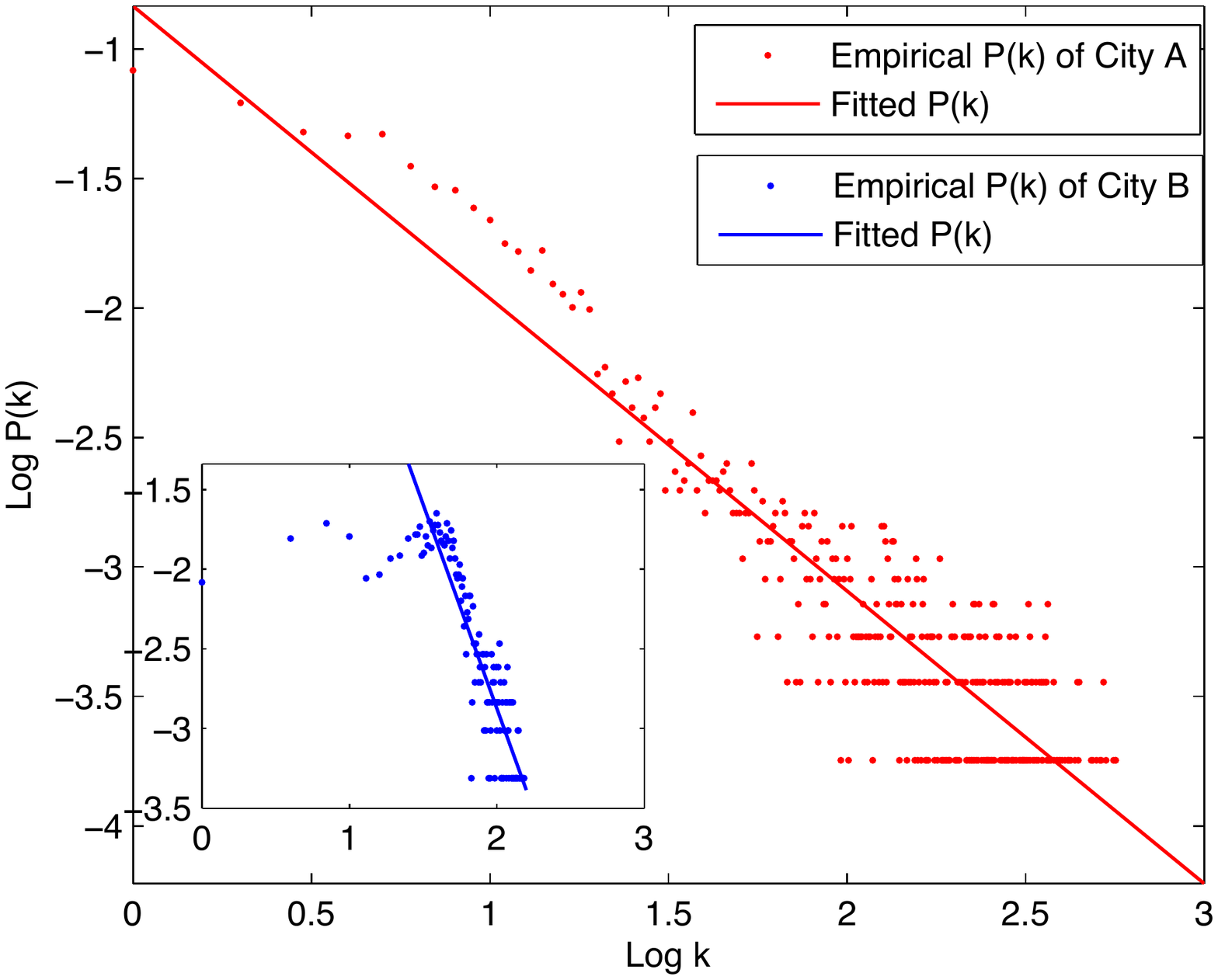}
	\caption{\footnotesize{Degree distributions of City A and City B with threshold being 0.54.}}
	\label{fig1}
\end{figure}

Apparently, the Pearson correlation coefficients vary along with the traffic similarity between BSs. For many trivial coefficients, the correlation between these two BSs could be neligible. Accordingly, we set a threshold $Z$ to evaluate the existence of one link between two BSs by comparing $\rho_{ij}$ and $Z$. Namely, if $\rho_{ij}$ is larger than $Z$, we think there exist a link between BS $i$ and BS $j$. In other words, we derive the adjacency matrix $W$ undirected graph from the calculated Pearson coefficient. After obtaining the graph, we observe that some BSs (nodes in the graph) are isolated from all the other BSs and have a zero degree. Therefore, we intentionally delete such nodes from the built graph.

\subsection{Threshold Selection}
In the process of modeling, choosing the appropriate threshold $Z$ is of great significance for our study. Thus, two aspects should be taken into consideration to choose the threshold $Z$. On one hand, $Z$ should be large enough, so as to avoid mistakenly assuming weakly correlated BSs to be linked. On the other hand, the choosen threshold needs to ensure the proportion of isolated BSs is relatively low for keeping a substantial graph size. In our work, we would like to process the datasets of the two cities with various threshold values $Z$, ranging from 0.5 to 0.7. Afterwards, we are going to analyze the relevant properties based on the spatial traffic correlation models we just established. As we expected, no matter how the threshold $Z$ changes, the properties of our model remain the same and more detailed information is shown in Table \uppercase\expandafter{\romannumeral2} and \uppercase\expandafter{\romannumeral3}. Without loss of generality, we fix the spatial traffic correlation model and study the targeted properties in Section \uppercase\expandafter{\romannumeral4} and Section \uppercase\expandafter{\romannumeral5} with the threshold $Z$ being equal to 0.54.   

 \begin{figure}
 	\centering
 	\includegraphics[width=0.5\textwidth]{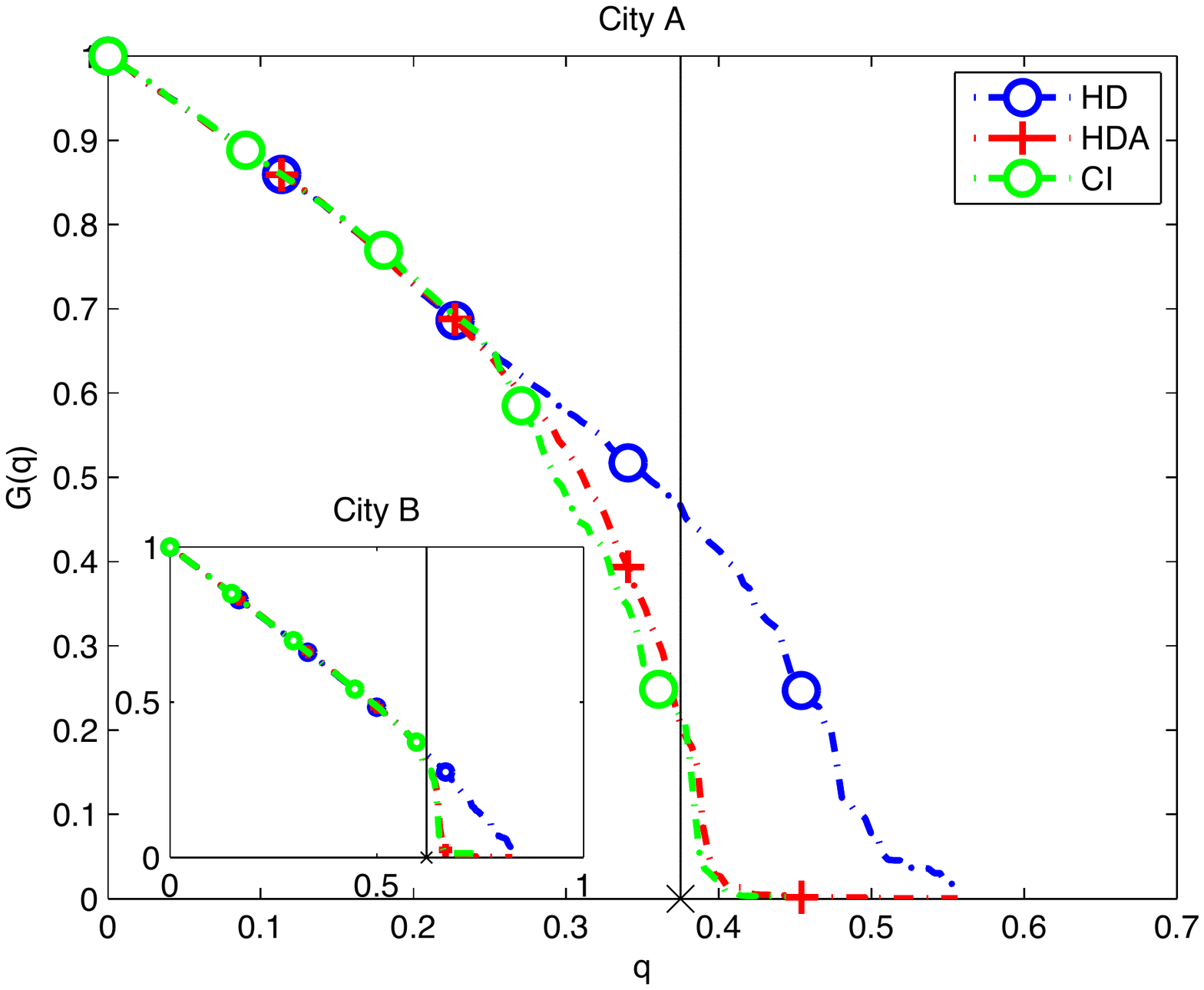}
 	\caption{\footnotesize{Performance of CI in correlation model compared with heuristic methods (HD, HDA).}}
 	\label{CI1}
 \end{figure}

\section{Analysis and Application of degree distribution}
\subsection{Degree Distribution}
As depicted in Section \uppercase\expandafter{\romannumeral3}, the spatial traffic correlation model is built in terms of the traffic loads and contributes to understanding the underlying relationship of BSs, which can not be directly observed from brief information such as locations (e.g., longitude and latitude) or BS types.

  \begin{figure}
  	\centering
  	\includegraphics[width=0.5\textwidth]{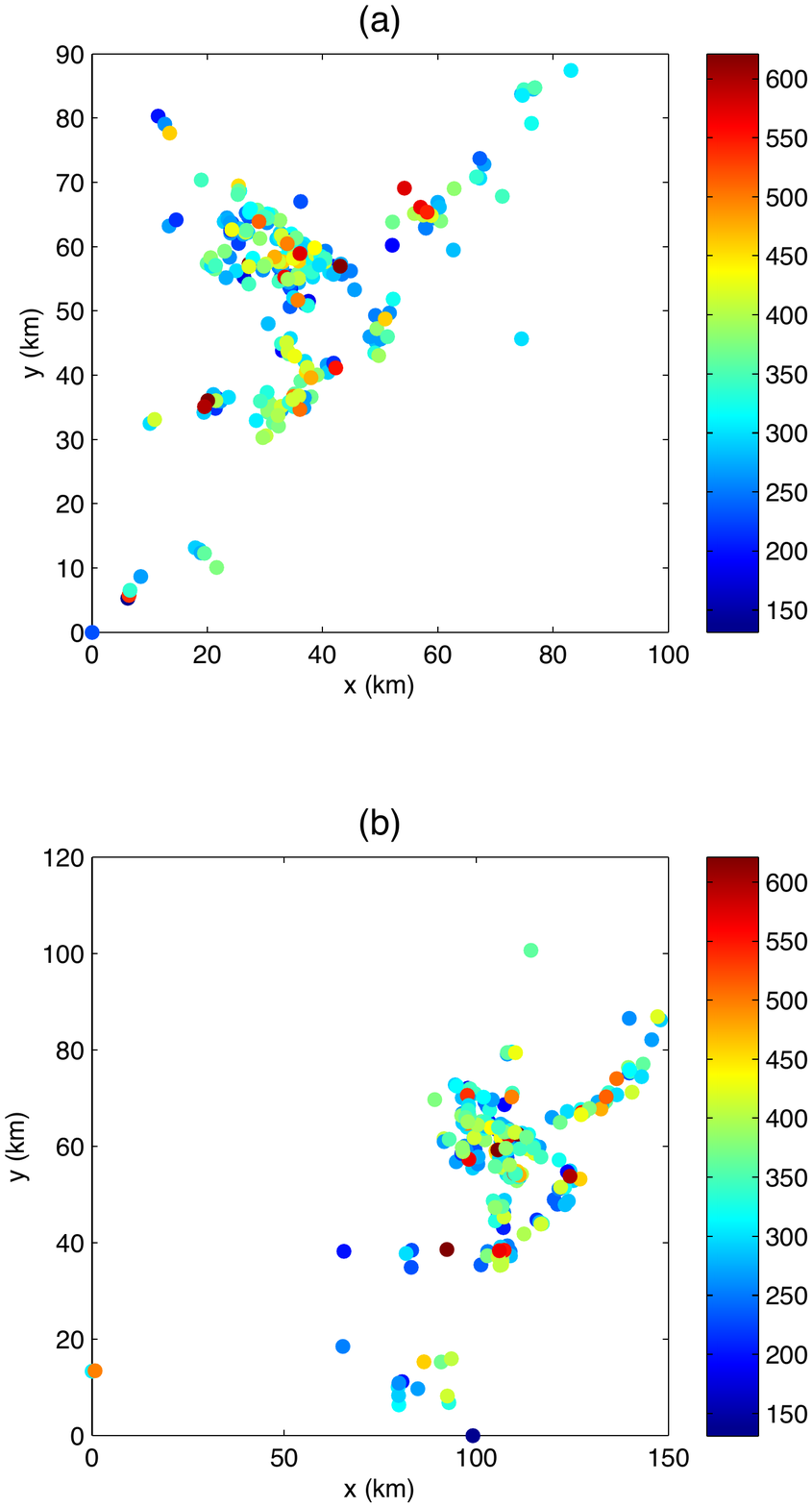}
  	\caption{\footnotesize{$a )$: The location of the most influential 250 base stations of City A is shown in the geographic space and the degree is color coded. $b )$: The counterpart of the remaining 250 base stations of City A.}}
  	\label{A_compare}
  \end{figure}
  
  \begin{figure}
  	\centering
  	\includegraphics[width=0.5\textwidth]{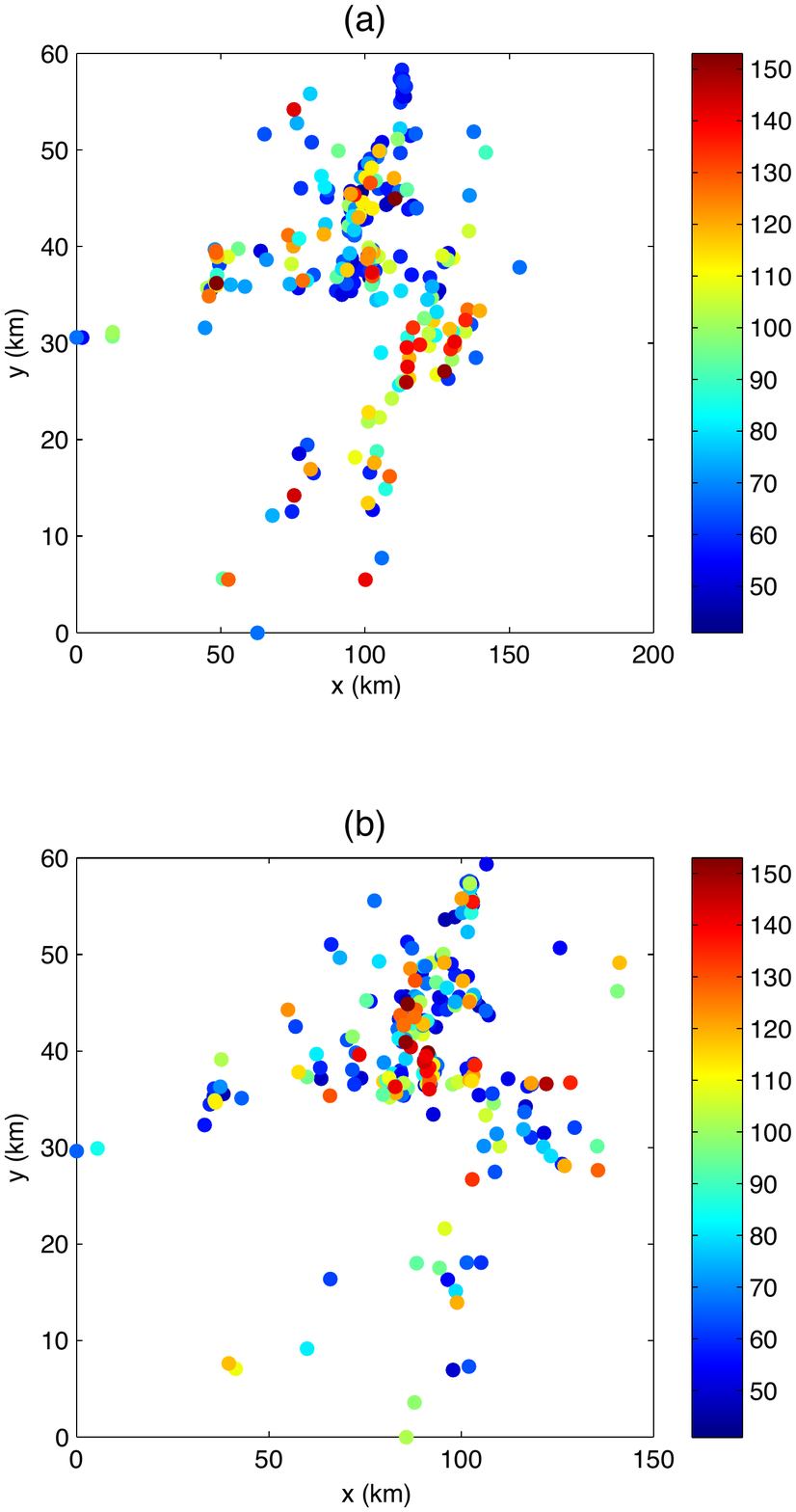}
  	\caption{\footnotesize{$a )$: The location of the most influential 250 base stations of City B is shown in the geographic space and the degree is color coded. $b )$: The counterpart of the remaining 250 base stations of City B.}}
  	\label{B_compare}
  \end{figure}

\begin{table}
	\newcommand{\tabincell}[2]{\begin{tabular}{@{}#1@{}}#2\end{tabular}}
	\centering
	\caption{Data reprocessing and degree Distribution analysis of Two Cities.}
	\setlength\abovecaptionskip{-5pt}
	\setlength\belowcaptionskip{-5pt}
	\scalebox{0.8}{
		\begin{tabular}{ccccccc}
			\toprule
			City & Threshold & \tabincell{c}{Network\\ Size $N$} & \tabincell{c}{Number of \\ Isolated BSs}& \tabincell{c}{Rate of\\ Isolated BSs} & \tabincell{c}{Degree \\ Exponent $\lambda$}\\
			\midrule
			\multirow {7}{*}{A} & 0.50 & 5046 & 527 & 9.46\% & 1.0394\\
			& 0.52  & 4800 & 773  & 13.87\%  & 1.0593\\
			& 0.54  & 4494 & 1079 & 19.36\%  & 1.1298\\
			& 0.56  & 4120 & 1453 & 26.07\%  & 1.1489\\
			& 0.58  & 3711 & 1862 & 33.41\%  & 1.2049\\
			& 0.60  & 3287 & 2286 & 41.02\%  & 1.2581 \\
			& 0.65  & 2228 & 3345 & 60.02\%  & 1.4088 \\
			\midrule
			\multirow {7}{*}{B} & 0.50 & 2050 & 3  & 0.15\% & 2.2790 \\
			& 0.54  & 2042& 11  & 0.54\% & 2.5951 \\
			& 0.56  & 2031& 22  & 1.07\% & 2.5955 \\
			& 0.58  & 2002& 51  & 2.48\% & 2.6544 \\
			& 0.60  & 1974& 79  & 3.85\% & 2.6191 \\
			& 0.65  & 1807& 246 & 11.98\% & 2.5635 \\
			& 0.70  & 1521& 532 & 25.91\% & 2.6483 \\
			\bottomrule
		\end{tabular}}%
		\label{tb:rmse}%
	\end{table}%

After setting up the spatial traffic correlation model, an adjacency matrix $W$ can be obtained while the definition of degree refers to the number of BSs that are highly correlated with a chosen BS. In other words, the degree of BS $i$ can be expressed by

\begin{equation}
	k_i=\sum_j w_{ij}
	\label{eq30}
\end{equation}

Meanwhile, $P(k)$ is defined as the probability that a randomly chosen vertex (BS) has degree $k$ \cite{Pastorsatorras2014Epidemic}. Since the number of BSs in City A is 5573, an adjacency square matrix with the dimension of 5573 can be obtained. It is verified that the obtained degree distributions are scale-free and obviously satisfy Power-law with almost constant exponents $\lambda$ for various thresholds $Z$ that are larger than 0.5 in Table \uppercase\expandafter{\romannumeral2}. Recalling the former statements in Section \uppercase\expandafter{\romannumeral3} and uniting the numerical results in Table \uppercase\expandafter{\romannumeral2}, although our conclusion remains the same under different thresholds $Z$, we advise that the suitable value of threshold $Z$ can be set within the range from 0.5 to 0.6 to make sure the proportion of removed BSs is acceptable. As illustrated in Fig. 2, we provide the fitting results of the degree distributions of City A and City B. In general, the spatial traffic correlation model points to the property of scale-free and help us to know which BSs have high degree values. The scale-free property from the traffic load correlation model clearly demonstrates that the minority of BSs with large degree are highly correlated with plenty of other BSs, while the other remaining BSs are only correlated with a few number of BSs. 

\begin{remark}
	Based on the spatial traffic correlation model, empirical and fitting results reveal that its degree distribution can be well depicted by a Power-law function. Specifically, few nodes are highly popular while most of the nodes are less popular in the network.
\end{remark}

\subsection{Identifying Influential BSs}
 Influential nodes usually play a decisive role on maintaining the network connectivity, enhancing network stability and improving the information transmission efficiency \cite{altarelli2013optimizing,kitsak2010identification,L2016The}. Similarly, the influential BSs can take more important roles in cellular networks. For example, cellular networks have already employed macrocell BS as the signaling node, so the macrocell BSs are more suitable to be influential nodes due to their greater coverage capability and being more easily to predict the tendency of BS traffic loads. As a result, it is imperative to pick out the most important BSs so as to assign them more functions such as signaling control. Given that it is critical to investigate how to find out the most influential BSs, we apply two heuristic strategies to identify the influential BSs considering our correlation model. Moreover, based on the theory of influence maximization in complex networks, we further employ the CI algorithm for localizing the most influential BSs \cite{morone2015influence,morone2016collective}.
 
Generally, the influential nodes are defined as a set of nodes, which is much smaller than the total network size, however, if removed, would break down the network into many disconnected components. At a general level, we use the size of the giant connected component to measure the remaining network structure when the influential nodes are removed from the network. In this paper, we aim at finding out the minimal set that guarantees a global connection of the network and the size of this minimal set $q_c$. The existence of the giant connected component can be expressed by $G(q)$ after removing a certain fraction $q$ of the network size. Then, our problem corresponds to finding the optimal set whose removal would dismantle the network:

\begin{equation}
	q_c=min\{q \in[0,1]:G(q)=0\}
	\label{eq30}
\end{equation} 

CI is an effective algorithm in terms of finding the most influential nodes, which removes nodes one-by-one according to their CI value:

\begin{equation}
CI_l(i)=(k_i-1) \sum_{j \in \partial Ball(i,l)} (k_j-1)
\label{eq30}
\end{equation} 
where $Ball(i,l)$ is the ball of radius $l$ centered on node $i$, and $\partial Ball(i,l)$ is the frontier of the ball, which is the set of nodes at distance $l$ from node $i$. CI algorithm removes the node with the highest $CI_l(i)$ value at each step, and the process is repeated until the giant component is destroyed \cite{morone2015influence}.

Fig. 3 shows the optimal threshold $q_c$ for the traffic load correlation model of City A and City B. In the same figure, we compare the optimal threshold against the other two heuristic methods: high-degree(HD), high-degree adaptive(HDA). For both cities, CI produces a smaller threshold, which represents a better performance of this algorithm. 

\begin{figure}
	\centering
	\includegraphics[width=0.5\textwidth]{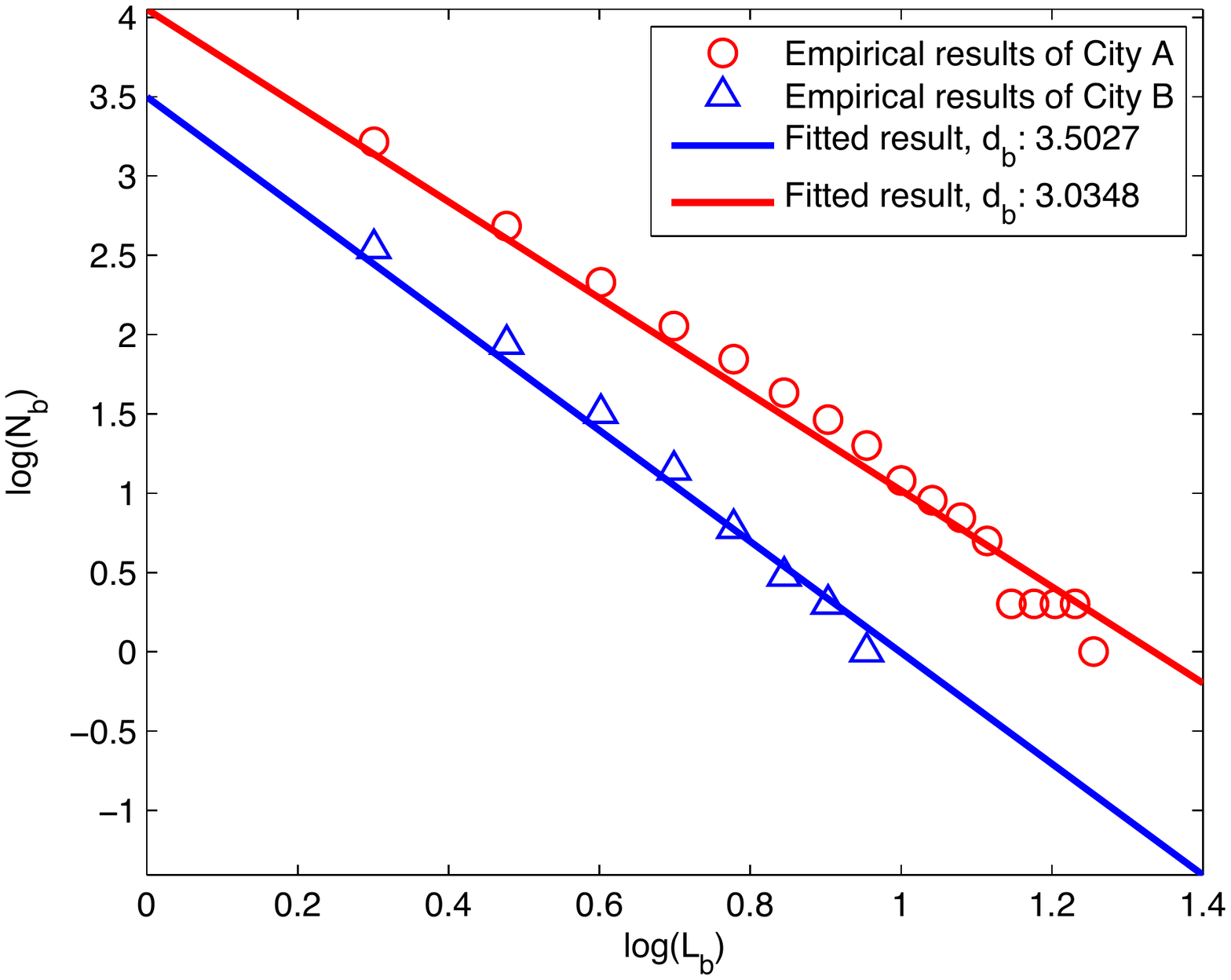}
	\caption{\footnotesize{Fractal patterns of City A and City B with the same threshold $K$ 0.54.}}
	\label{fractal}
\end{figure}   

Afterwards, according to the optimal set of nodes found by the CI algorithm, we display the locations of the most influential 500 base stations of City A in the map and color codes each BS's degree in Fig. 4 and Fig. 5. From the two figures, we observe that among the most influential base stations extracted by the CI algorithm, a large number of low-degree BSs even exhibit a greater influence than some high-degree BSs. That is to say, we should pay more attention to those influential BSs even with low-degree, comparing with the high-degree BSs with less influence.

\begin{remark}
	According to numerical results, degree is not always a better criteria in measuring the node influence. Namely, a number of low-degree BSs appear to be more significant than some BSs with larger degree values.
\end{remark}

\section{Structural Properties of The Traffic Load Correlation Model}
\subsection{Fractal Patterns}
One important property that exists in many complex and real-world networks is fractality \cite{Gallos2007A}. In fractal geometry \cite{mandelbrot1983fractal}, box-covering \cite{Song2005Self} is widely used to approximately evaluate the fractal dimension of a fractal object. Based on this method, fractal networks can be characterized by the following scaling relations:
\begin{equation}
	N_b(L_b)/N \sim L_{b}^{-d_b}
	\label{eq30}
\end{equation}
where $L_b$ denotes the size of boxes used to cover the network and $N_b(L_b)$ is the minimum number of boxes among all the possible tiling configurations with the box size equaling to $L_b$. Accordingly, the fractal dimension can be calculated through the following equation:

\begin{equation}
	d_b \sim \lim_{L_b \rightarrow 0} \frac{\log N_b(L_b)}{-\log L_b}
	\label{eq30}
\end{equation}

In reality, the value of $d_b$ can be obtained by fitting the slope between log$N_b$($L_b$) and log $L_b$. After these early-stage preparations, we employ the box-covering method to investigate the traffic load correlation model of BSs. In this paper, we carry out the fractal pattern analysis with the size of box varying from 1 to the diameter of network, namely, 17 for City A and 7 for City B. This means that when $L_b$ is no less than the network diameter, the value of $N_b$ must be 1. Fig. 5 shows the results from the box-covering algorithm applied in City A and City B, respectively.

As illustrated in Fig. 6, for City A, the relation between log($N_b$) and log($L_b$) can be well-fitted by a straight line, which implies a clear fractal property of the network. Moreover, the fractal dimension $d_b$ approximates 3.0348 with the $R$ square value being 0.9460 denoting the good fitness of the curve. Meanwhile, for City B, the fractal dimension approaches 3.5027 with the $R$ square value being 0.9532.

\subsection{Skeleton Features}
Regardless of the entanglement, a network always possesses a "skeleton" to simply represent the network structure and understand the topological organization \cite{Goh2006Skeleton}. The skeleton is a particular spanning tree consisting of edges with the highest betweenness centralities \cite{Kitsak2007Betweenness}. Plenty of researches have elaborated the importance of skeleton in understanding the topological organization of a complex network.

\begin{figure}
	\centering
	\includegraphics[width=0.5\textwidth]{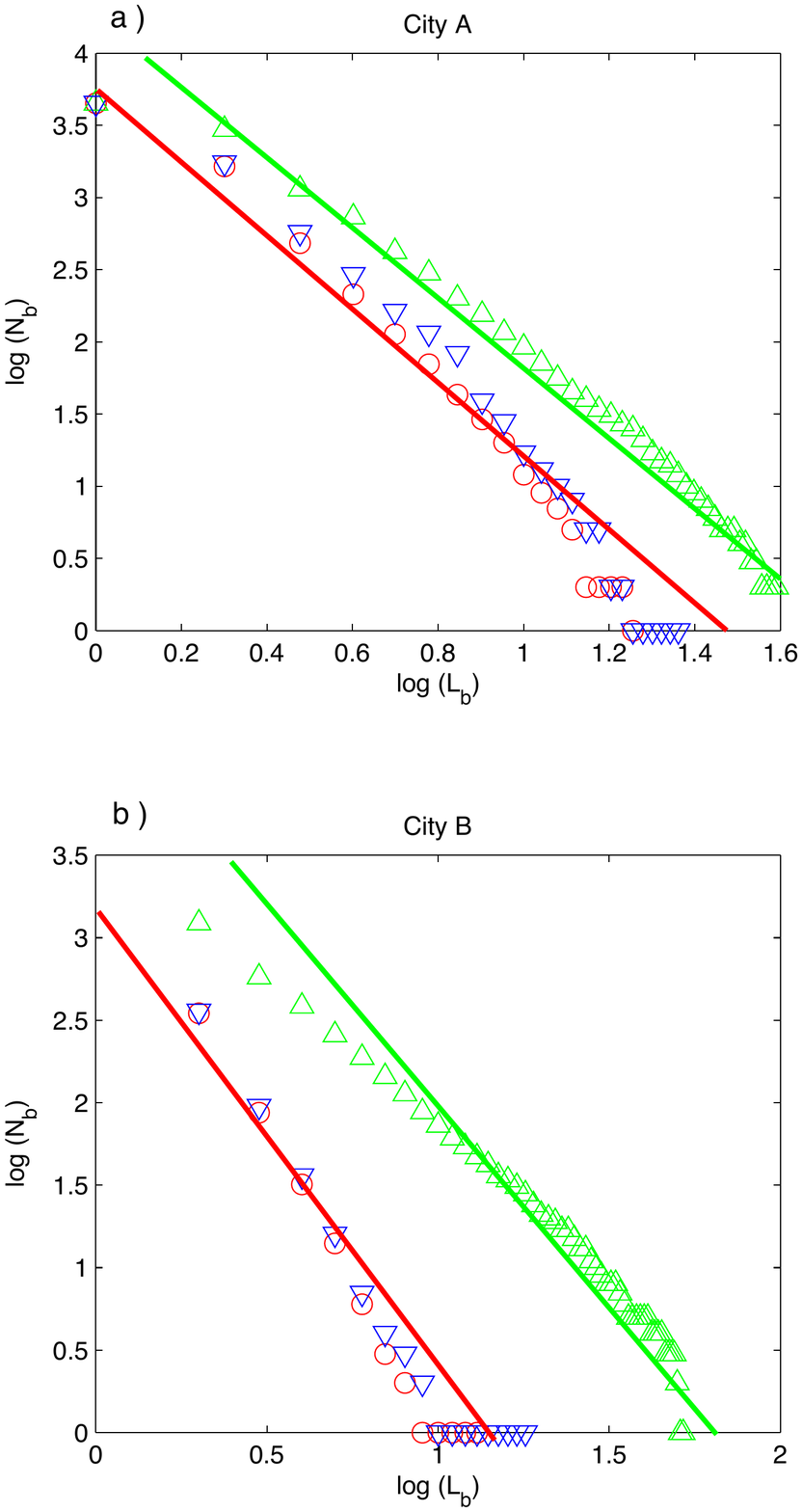}
	\caption{\footnotesize{Box-covering analysis of the original network ({\color{red}{$\circ$}}), its skeleton ({\color{blue}{$\triangledown$}}) and random spanning tree ({\color{green}{$\triangle$}}) of City A and City B}}
	\label{skes}
\end{figure}

Basically, skeleton is thought to be a maximum spanning tree. Thus, the skeleton of our correlated BSs network is a spanning tree connected by the most close links, whose topology can be regarded as the core of the correlated BSs network. Inspired by the classical Prim and Kruskal algorithms for building the minimum spanning tree, we propose a modified algorithm to find out the skeleton of the traffic load correlation model (i.e. Algorithm 2). 

\begin{algorithm}
	\caption{Modified algorithm used to extract skeleton }
	\label{alg:A}
	\begin{algorithmic}
		\STATE input: $G=(V, E)$, adjacency matrix $W$ ;
		
		 betweenness centralities matrix EC;
		\STATE output:$P, Q$;
		\STATE initialization:
		\STATE {set $P=\{v_1\}, Q=\emptyset$.} 
		\REPEAT 
		\STATE find the maximum value EC$(p, v)$, $p \in P, v \in V$;
		\STATE set ${P=P+\{v\}}$;
		\STATE set ${Q=Q+\{pv\}}$;  
		\UNTIL ($P=V$)
	\end{algorithmic}
\end{algorithm} 

Following Algorithm 2, we extract the skeletons for the spatial traffic correlation models and study their degree distribution along with fractality. Numerical results verify that the skeletons are also scale-free with exponent values $\lambda$ equaling 2.214 and 2.152 for City A and B. Furthermore, after tiling the skeletons with the box-covering algorithm, the number of boxes needed to cover the networks is almost identical with the original networks. The box-covering analysis results of the original network, the skeleton and the random spanning tree are provided in Fig. 7. According to the curves, the relevant results express that although the random spanning tree possesses a different statistics of $N_b$, the fractal dimensions of the random spanning tree and the original network are just the same. Meanwhile, the fractality of the skeleton matches the fractality of the original correlation model very well. Hence, understanding the properties of the skeleton is of great importance for analyzing the original model. 

\begin{remark}
	Detailed analyses explain the special structure (i.e., fractal patterns) of spatial traffic correlation model. In the meantime, its skeleton, which exhibits identical features, is of great contribution for analyzing the original network. 
\end{remark}

\subsection{Further Exploration on Small-World}
The small-world property usually coexists with scale-free networks \cite{Li2015A}. Specifically, small-world property refers to the average distance $d$ scales logarithmically with the network size $N$ as $d$$\sim$$\ln N$. Another indispensable characteristic of small-world networks is their high clustering coefficient \cite{Pastorsatorras2014Epidemic}. Structural analysis of the traffic load correlation model tells us that the size is 4494 for City A while its average distance $d$ equals to 4.5257. The relationship between $d$ and $N$ conforms to the above equation. City B with size 2042 and $d$ equaling to 3.3947 also meets this mathematical expression. Moreover, the clustering coefficients can also be obtained, being equal to 0.5144 and 0.5177, respectively, which represents a highly clustering feature. We do ensure that the spatial traffic correlation model of BSs possesses the small-world property and more supporting evidences are given in the following as well as in Table  \uppercase\expandafter{\romannumeral3}.

\begin{enumerate}
	
\item Pearson Coefficient

Degree of assortativity is one of the important features to describe network. Degree-degree correlations can be characterized by Pearson coefficient, which is defined as:
	
\begin{equation}
	r=\frac{M^{-1} \sum_{e_{ij}} k_i k_j-[M^{-1} \sum_{e_{ij}} \frac{1}{2} (k_i+k_j)]^2}{M^{-1} \sum_{e_{ij}} \frac{1}{2} (k_i^2+k_j^2)-[M^{-1} \sum_{e_{ij}} \frac{1}{2} (k_i+k_j)]^2}
	\label{eq30}
\end{equation}

\begin{figure}
	\centering
	\includegraphics[width=0.5\textwidth]{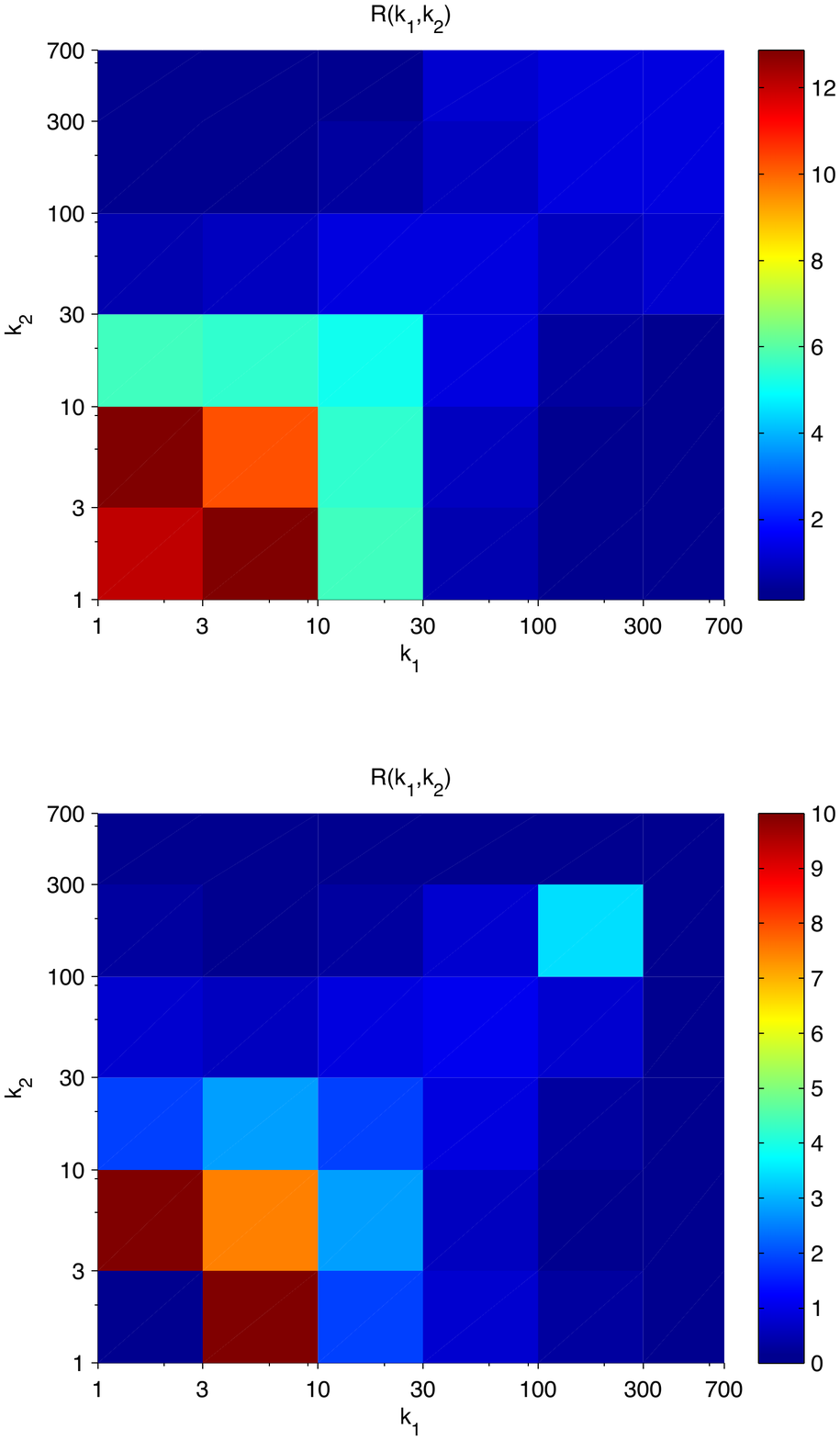}
	\caption{\footnotesize{ $a )$: Correlation profile to compare the hub-hub correlation emerging from the correlation model of City A. $b )$: Correlation profile to compare the hub-hub correlation emerging from the correlation model of City B.}}
	\label{compare1}
\end{figure}

where $M$ denotes the total number of edges,$k_i$ and $k_j$ are the degrees of the two vertices at the ends of edge $e_{ij}$. The Pearson coefficient $r$ ranges from -1 to 1, being positive for assortative networks and negative for disassortative ones. The Pearson coefficients of our models are 0.1535 and 0.5362 for City A and City B, respectively. In other words, the spatial traffic correlation models are assortative \cite{yook2005self}.

\item Correlation Profile
	
Correlation profile is a metric of great importance to explain the structural information and the statistical property of correlation between the nodes within a network configuration  \cite{L2016The}. The correlation profile is defined as:
	
\begin{equation}
	R(k_1,k_2)=P(k_1,k_2)/ P_r(k_1,k_2)
	\label{eq30}	
\end{equation} 
	
where $P(k_1,k_2)$ is the joint probability distribution representing the probability of finding a node with $k_1$ links connected to a node with $k_2$ links. While $P_r(k_1,k_2)$ is acquired by randomly swapping of the links with the degree distribution remaining unchanged. The plot of the ratio $R(k_1,k_2)$ demonstrates a correlated structure that deviates from the random uncorrelated case. We apply this metric to depict the correlation models and the corresponding results are shown in Fig. 8.
\end{enumerate}

\begin{table}
	\newcommand{\tabincell}[2]{\begin{tabular}{@{}#1@{}}#2\end{tabular}}
	\centering
	\caption{Structure and Property Analysis Of Two Cities.}
	\setlength\abovecaptionskip{-5pt}
	\setlength\belowcaptionskip{-5pt}
	\scalebox{0.8}{
		\begin{tabular}{ccccccc}
			\toprule
			City & Threshold  & \tabincell{c}{fractal\\ dimension $d_b$ } & \tabincell{c}{Network\\ Size $N$} & \tabincell{c}{Average\\ Distance $d$} & \tabincell{c}{Clustering \\ Coefficient}& \tabincell{c}{Pearson\\ Coefficient $r$}\\
			\midrule
			\multirow {6}{*}{A} & 0.50 & 3.7944 & 5046 & 3.7944 & 0.5146 & 0.1461 \\
			& 0.54 & 3.0348 & 4494 & 4.5257 & 0.5144 & 0.1535 \\
			& 0.56 & 3.5422 & 4120 & 4.9610 & 0.5094 & 0.1577 \\
			& 0.58 & 2.7239 & 3711 & 5.5517 & 0.5009 & 0.1603 \\
			& 0.60 & 2.4134 & 3287 & 6.5524 & 0.4843 & 0.1761 \\
			& 0.65 & 2.3755 & 2228 & 6.8535 & 0.5214 & 0.1593 \\
			
			\midrule
			\multirow {6}{*}{B} & 0.50 & 3.7120 & 2050 & 2.9981 & 0.4972 & 0.5378 \\
			& 0.54 & 3.5027 & 2042 & 3.3947 & 0.5177 & 0.5362 \\
			& 0.58 & 3.1010 & 2002 & 3.9098 & 0.5342 & 0.5291 \\
			& 0.60 & 3.0007 & 1974 & 4.2315 & 0.5378 & 0.5221 \\
			& 0.65 & 3.0286 & 1807 & 5.2704 & 0.5470 & 0.5143 \\
			& 0.70 & 3.3045 & 1521 & 6.5584 & 0.5439 & 0.5042 \\
			\bottomrule
		\end{tabular}}%
		\label{tb:rmse}%
	\end{table}%

From Fig. 8, we observe that the models exhibit a higher degree of correlation, namely, nodes with large degree tend to be connected with nodes of large degree and vice versa, which is the primary cause that contributes to the small-world behavior. While the emergence of scale-free fractal networks is due to the repulsion between nodes of large degree, fractality seems to be contradicted with small-world phenomenon. Nevertheless, empirical results suggest that there exist networks with the simultaneous appearance of both fractal and small-world properties, for which a mathematical generation model has been given in \cite{Song2005ARTICLES}. 

We have demonstrated that the spatial traffic correlation model of BSs expresses scale-free, fractal and small-world properties simultaneously, which will further facilitate the performance analysis of complex cellular networks as well as the design of efficient networking protocols. Firstly, scale-free behavior signifies the heterogeneous network structure. In particular, the minority of BSs are correlated with a large number of BSs, which may play pivot roles in cellular networks. Secondly, fractality explains the possibility that the degree distribution might remain unchanged under scale transformation and leads to network self-similarity. Moreover, for a topological structure with fractality, we can find some regualrities from its special topology and irregularity, which contibutes to more effective resource assignment based on dynamic BSs management. Finally, the discovery of small-world property means that, despite the large-scale feature of the traffic load correlation model, the traffic association on base stations is very compact.

\section{Conclusions}
In this paper, we have proposed a unique approach to establish the spatial traffic correlation model for the base stations in complex cellular networks, leveraging a traffic load vector with the elements being the traffic data crossing each BS in a certain interval. We first created the spatial load correlation model according to the Pearson coefficient values between various BSs. Afterwards, based on the correlation model, we discovered that the spatial correlation structure is scale-free along with the coexistence of fractality and small-world feature, after careful verification in terms of common metrics in the literature. Additionally, we extracted the skeleton of the spatial correlation model in order to search for the most compact pairs of BSs to obtain the most significant links in our model. Moreover, we conducted some comparisons between CI algorithm and two best heuristic methods to pick out the set of the most influential base stations. Finally, several suggestions on the potential main applications in real networking scenarios were provided in Section \uppercase\expandafter{\romannumeral4}.\\

\bibliographystyle{IEEEtran}
\bibliography{edition1}

\end{document}